\documentclass[prl,floatfix,twocolumn,amsmath]{revtex4}
\usepackage{graphicx}
\usepackage{color}

\begin{document}

\def\ket#1{|#1\rangle}
\def\bra#1{\langle#1|}
\def\av#1{\langle#1\rangle}
\def\myarrow{\mathop{\longrightarrow}}
\def\ua{\uparrow}
\def\da{\downarrow}
\def\dpt#1{#1\rangle}
\setlength\abovedisplayskip{6pt}
\setlength\belowdisplayskip{6pt}
\setlength\belowcaptionskip{-8pt}
\setlength{\mathindent}{0pt}


\title{Optical control of Feshbach resonances in Fermi gases using molecular dark states}

\author{Haibin Wu $^{1,2}$ and J. E. Thomas$^{2}$}
\affiliation{$^{1}$Department of Physics, Duke University, Durham, NC 27708, USA}
\affiliation{$^{2}$Department of  Physics, North Carolina State University, Raleigh, NC 27695, USA}

\date{\today}

\begin{abstract}
We propose a general method for optical control of  magnetic Feshbach resonances in ultracold atomic gases with more than one  molecular state in an energetically closed channel. Using two optical frequencies to couple two  states in the closed channel, inelastic loss arising from spontaneous emission is greatly suppressed by destructive quantum interference at the two-photon resonance, i.e., dark-state formation, while the scattering length is widely tunable by varying the frequencies and/or intensities of the optical fields.  This technique is of  particular interest for a two-component atomic Fermi gas, which is stable near a Feshbach resonance.
\end{abstract}
\maketitle


Ultracold atomic gases with controllable interactions are now widely studied by exploiting collisional (Feshbach) resonances~\cite{Reviews}.  In contrast to Bose gases, which suffer from three-body inelastic processes near a resonance, two-component Fermi gas mixtures are stable as a result of the Pauli principle,  and can be rapidly cooled to quantum degeneracy by evaporation in the resonant regime~\cite{OHaraScience}. Typically, in a Feshbach resonance, an external magnetic field controls the interaction strength between spin-up and spin-down atoms, by tuning the  energy of an incoming, colliding atom pair  into resonance with that of a bound molecular state in an energetically closed channel~\cite{Stoof, CHIN}. Optical tuning methods offer advantages over magnetic tuning, such as  rapid temporal control and high resolution spatial control of the interaction strength near a Feshbach resonance, opening many new fields of study, such as nonequilibrium strongly interacting Fermi gases~\cite{Bulgac}.
The use of electromagnetically induced transparency (EIT) to control Feshbach resonances was suggested by Harris~\cite{HarrisFB}. Optical control of Feshbach resonances has been explored previously in Bose gases~\cite{GrimmFB, PLett} and currently is receiving substantial attention~\cite{rempe}. Optical Feshbach resonances (OFR), which employ photoassociation light to drive a transition from the continuum of the incoming atom pair state to an excited molecular bound state, has been proposed and experimentally observed~\cite{walraven, Julienne, PLett, Theis, GrimmOFB, enomoto}. However, light-induced inelastic collisions and the accompanying loss limit its practical applicability. Submicron-scale spatial modulation of an inter-atomic interaction has been observed in an alkaline-earth atomic condensate~\cite{yamazaki}.  Recently, Rempe and coworkers have used a single optical field to control the scattering length near a magnetic Feshbach resonance  by driving a transition between a ground state  in the closed channel and an excited molecular state. In this method, a large laser intensity  and a large frequency detuning are required for suppressing the light-induced loss~\cite{rempe}. OFR also has been studied by using a narrow intercombination line of  a bosonic gas $^{88}$Sr, with the laser frequency tuned far away from  resonance~\cite{Ye}. Unfortunately, all of these methods suffer from limited tunability of the scattering length as well as loss and heating, which arise from light-induced inelastic collisions.

In this Letter, we suggest a general ``dark-state" optical method  for widely controlling the interaction strength near a magnetic Feshbach resonance, while suppressing spontaneous scattering by quantum interference, in ultracold atomic gases with at least two molecular states in the closed channel. In a  Fermi gas near a broad Feshbach resonance, this method yields a {\it double} suppression of the spontaneous scattering rate, as the probability of occupying the closed channel molecular state near and above resonance has been measured to be very small, $\leq 10^{-5}$ in $^6$Li~\cite{Hulet}.

\begin{figure}[htb]
\includegraphics[width=2.75in]{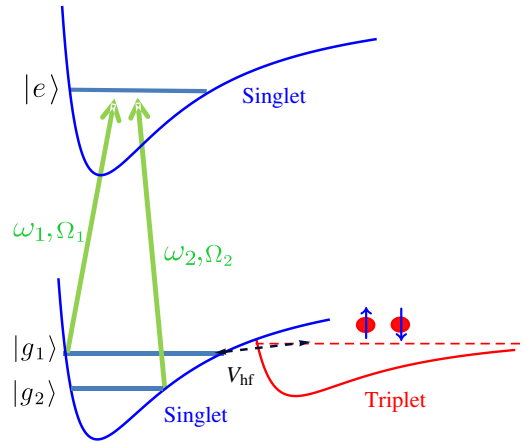}
\caption[example]
   { \label{fig:fig1}
Scheme for ``dark-state" optical control of a Feshbach resonance using two closed-channel molecular states. Optical fields of frequencies $\omega_{1}$ and $\omega_{2}$ and Rabi frequencies $\Omega_{1}$ and $\Omega_{2}$, respectively,  couple ground singlet molecular states $\ket {g_{1}}$ and $\ket {g_{2}}$ to the excited molecular state $\ket e$;  $V_{hf}$ is the hyperfine  coupling $V_{hf}$ between the incoming atomic pair state  in the open (triplet) channel and $\ket {g_{1}}$, which is responsible for a magnetically controlled Feshbach resonance. }
\end{figure}

The basic scheme, Fig. 1,  is illustrated for a pair of atoms in two hyperfine states (denoted spin-up and spin-down), which undergoes an s-wave collision in the ground electronic state triplet molecular potential (open channel). The hyperfine interaction couples the scattering continuum of the open channel to a bound singlet vibrational state $\ket {g_{1}}$ in the closed channel. An applied bias magnetic field $B$ tunes the total energy of the colliding atom pair downward, near $\ket{g_1}$, producing a collisional (Feshbach) resonance. A second molecular ground state $\ket {g_{2}}$ is not coupled to the open channel; For example, $\ket {g_{2}}$ can be a different singlet vibrational state. Two optical fields with frequencies $\omega_{1}$  and $\omega_{2}$ couple $\ket {g_{1}}$ and  $\ket {g_{2}}$ to the electronically excited singlet vibrational state $\ket e$. To determine the s-wave scattering length in the presence of the light fields, we use a method similar to that employed by Fano~\cite{Fano}.

We write the Hamiltonian as
\begin{align}
H=&H'_{hf}-\mu_z B+E_{g_1}\ket {S_1, g_1} \bra {S_1, g_1}+E_{g_2}\ket {S_2, g_2} \bra {S_2, g_2} \nonumber\\
&+E_e\ket {S_e, e} \bra {S_e, e}+\frac{p^2}{m}\ket {T, k}\bra {T, k}+H_{int}
\label{eq:Hamiltoniandensity}
\end{align}
where the optical interaction Hamiltonian is
\begin{align}
H_{int}=- \hbar \Omega_{1}  \cos (\omega_1 t)\ket {g_{1}}\bra {e}-\hbar \Omega_{2}\cos(\omega_2t)\ket {g_{2}}\bra e+h.c. \nonumber
\end{align}
  Here, $E_{j}$ ($j=g_1, g_2, e$)  is the molecular internal energy and $\mu_z$ the magnetic dipole moment operator.  $H'_{hf}$ is the hyperfine interaction with matrix elements between the triplet $T$ and singlet states $S_{1,2}$ given by $\bra {S_{1}} H'_{hf}\ket T=V_{hf}$ and $\bra {S_{2}} H'_{hf}\ket T=0$. The energy of the triplet state is  magnetic field dependent, $\bra T (H_{hf}'-\mu_z B)\ket T\equiv E_T$. $\Omega_{1,2}=\frac{\bra e d\ket {g_{1},g_{2}}{\cal E}_{1,2}}{\hbar}$ are the Rabi frequencies corresponding to the dimer transitions $\ket {g_{1}}\rightarrow\ket e$ and $\ket {g_{2}}\rightarrow\ket e$, respectively.

The time-dependent wavefunction  takes the form
\begin{align}
\ket{\psi_E(t)} =c_1\ket {S_{1}, g_1}+c_2 \ket {S_{2} , g_2}+c_e \ket {S_e, e}\nonumber\\
\int_{k'\neq k}d^3k' c_{T}(k')\ket {T, k'}+\tilde{c}_{T}(k)\ket {T ,k}. \label{eigenstate}
\end{align}
 Here $\tilde{c}_{T}(k)\ket {T ,k}$ represents the chosen incoming state with energy $E=E_T+\hbar^2k^2/m$, where $k$  is  the wavevector for the relative momentum of the colliding atoms.

 We take   $c_j=b_je^{-i[E+\hbar\omega_1\delta_{ej}-\hbar(\omega_2-\omega_1)\delta_{2j}]t/\hbar}$ (where $j=1,2,e, T$). Using the rotating wave approximation and assuming that the amplitudes $b_j$ of  the molecular states $\ket {g_{1}}$, $\ket {g_{2}}$ and $\ket e$ vary slowly compared compared to $\gamma_e$ and $|\Omega_1|^2/(\Delta_{e}-i\gamma_e/2)$, we obtain
\begin{align}
0=& \, \Delta_{g} b_1-\frac{ \Omega_1^*}{2}b_e+\!\!\!\!\!\int\limits_{k'\neq k }\!\!\!\!\! d^3k' g(k')b_T(k')+ g(k)\tilde{b}_T(k) \tag{3.a}\label{equation31}\\
0=&\, \delta b_2-\frac{\Omega_2^*}{2}b_e \tag{3.b} \label{equation32}\\
0=&\, -(\Delta_e+\frac{i\gamma_e}{2})b_e-\frac{\Omega_1}{2}b_1-\frac{\Omega_2}{2}b_2 \label{equation33}\tag{3.c}\\
0=&\, b_T(k')\,(E_T+\frac{\hbar^2k'^2}{m}-E)/\hbar +g(k')b_1 ,\label{equationTS}\tag{3.d}
\end{align}
where $g(k)\equiv V_{hf}\bra k\dpt{g_1}/\hbar$.  Here, $\Delta_{g}=(E_{g_1}-E)/\hbar$ and the single photon detuning is $\Delta_{e}=\omega_{1}-(E_e-E)/\hbar$.  The two-photon detuning is $\delta\equiv(\omega_2-\omega_1)-(E-E_{g_2})/\hbar$. The radiative decay rate of the molecular excited state is $\gamma_{e}=2/\tau_{spont}$, where $\tau_{spont}$ is the atom spontaneous lifetime. When the hyperfine coupling between $\ket{g_{1}}$ with the atomic pairs is strong,  as occurs near  a broad s-wave Feshbach resonance, adiabatic conditions may be difficult to achieve for bosons, which suffer from three body collisional loss, but are readily achieved for fermions.

\begin{figure}[htb]
\includegraphics[width=3.0in]{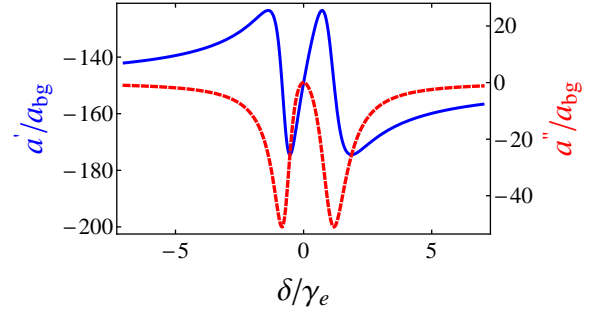}
  \caption{Scattering length as a function of the effective one-photon detuning  $\delta$ in units of $\gamma_{e}$. Real $a^{'}/a_{bg}$ (Blue  curve) and imaginary $a^{''}/a_{bg}$ (Red curve).  The parameters used here are for $^6$Li: $\Delta B=300$ G,  $\Delta\mu/\hbar=2\mu_B/\hbar=2\pi\times 2.8$ MHz/G, $ \gamma_{e}=2\pi \times 12$ MHz, and $a_{bg}=-1405\,a_0$; We take $\Omega_{1}=0.8 \gamma_{e}$, $\Omega_{2}=2 \gamma_{e}$, $\omega_2=\omega_{eg_{2}}$, $B-B_{0}= 2$ G. }
\label{fig:fig2}
\end{figure}

The  steady state solutions for the molecular amplitudes determine $b_{e}$ in terms of $b_1$, which then determines the ratio  $b_1/\tilde{b}_T$.  Eq.~(\ref{equationTS}) yields
\begin{align}
b_T(k')=\frac{g(k')}{E-E_T-\frac{\hbar^2k'^2}{m}}b_1 .\tag{4}\label{tripletwithsinglet}
\end{align}
When the interatom distance $r$ is large,  the molecular wavefunctions $\rightarrow 0$ and the continuum state $\ket k$ takes its asymptotic form $\propto\sin(kr+\delta_{bg})/kr$, where $\tan\delta_{bg}= - k a_{bg}$ is the phase shift arising from the background scattering in the triplet potential.   Inserting eq.~(\ref{tripletwithsinglet}) into eq.~(\ref{eigenstate}), the scattering state $\ket{\psi_{sc}}$ is given by
\begin{align}
\ket{\psi_{sc} (r\rightarrow \infty)}=\sin(kr+\delta_{bg}+\delta_{res})\frac{ \tilde{b}_T}{r}\ket T , \tag{5}
\end{align}
where $\delta_{res}=\arctan(\frac{2 \pi^2 m kg(k)}{\hbar^2}\frac{b_1}{\tilde{b}_T})$ is the phase shift arising from the coupling of the incoming continuum atoms state and the vibrational states of the singlet potentials.

Using the definition of the scattering length, $a=-\lim_{k\rightarrow 0}(\frac{\delta_{bg}+\delta_{res}}{k})$, the complex-valued s-wave scattering length, $a=a'+ia''$, takes the simple form

\begin{widetext}
\begin{align}
a=a_{bg}+\frac{2\pi^2 g(k)^2m/\hbar^2}{E-E_{g_1}-\!\!\!\int\limits_{k'\neq k }\!\!\! d^3k'\frac{g(k')^2}{E-E_T-\frac{\hbar^2k'^2}{m}}+\frac{\hbar\delta\Omega_1^2/4 }{(\Delta_e-i\gamma_e/2)\delta-\Omega_2^2/4 }}\Bigg|_{k\rightarrow 0}  \tag{6}. \label{scatteringlength}
\end{align}
\end{widetext}

For $\Omega_{1}=0$, Eq.~\ref{scatteringlength} immediately yields the well-known result for a magnetically induced
Feshbach resonance
\begin{align}
a=a_{bg}\bigg(1-\frac{\Delta B}{B-B_{0}}\bigg). \tag{7}
\end{align}
Here  the energy detuning is $-\Delta \mu(B-B_0)=E-E_{g_1}-\Delta E(k\rightarrow 0)$, where  $\Delta \mu$ is the difference between the magnetic moments of an atom pair and a molecule in state $g_1$ and $B_0$ is the resonant magnetic field. The energy shift arising from the hyperfine coupling of $\ket{g_1}$ to the continuum is $\Delta E(k)\equiv\int\limits_{k'\neq k }\, d^3k'\frac{g(k')^2}{E-E_T-\frac{\hbar^2k'^2}{m}}$ and $\Delta B=\frac{2\pi^2 g(k)^2m/\hbar^2}{a_{bg}\Delta\mu}|_{k\rightarrow 0}$ is the resonance width. In general,
\begin{align}
a'=a_{bg}\bigg(1-\beta\frac{\Gamma_2 (4\Delta_0\Gamma_2-\Omega_1^2\delta)+(\gamma_e\delta)^2\Delta_0}{4(\Delta_0\Gamma_2-\delta|\Omega_{1}|^{2}/4)^{2}
 +(\Delta_0\delta\gamma_{e})^{2}}\bigg),\tag{8}
 \label{eq:reala}
\end{align}
where $\beta=\Delta B\Delta\mu/\hbar$ and $\Gamma_2=\Delta_{e}\delta-\Omega_{2}|^{2}/4$.

The two body loss rate constant arising from optical scattering, $K_2(m^3/s)=-8\pi\hbar\,a^{''}/m$ is
\begin{align}
K_{2}=\frac{\gamma_{e}\delta^{2}|\Omega_{1}|^{2}\alpha}{4(\Delta_0\Gamma_2-\delta|\Omega_{1}|^{2}/4)^{2}
 +(\Delta_0\delta\gamma_{e})^{2}}, \tag{9}
 \end{align}
 where $\alpha=4\pi a_{bg}\Delta B\Delta \mu/m$. We see that $K_2$ is suppressed by the square of the two photon detuning $\delta$ for $\Delta_0=\Delta\mu(B-B_{0})/\hbar\neq0$. Actually, the $\Omega_1$ field also weakly drives transitions of the incoming atom pair states (mostly triplet) to the excited molecular state $\ket e$ (singlet), with a small Rabi frequency $f\Omega_{1}$, where $f<<1$. Near the Feshbach resonance, where $\Delta_0$ is small, we find that the corresponding photoassociation rate is suppressed.  Further, one can show that even including photoassociation, the two body loss is still proportional to $\delta^2$  and is  thus greatly suppressed.  Interference in the $f$-dependent term  causes a minimum in the photoassociation rate~\cite{Mackie}, which we will discuss in more detail elsewhere.

The s-wave scattering length and corresponding loss as a function of the effective two-photon detuning  $\delta/\gamma_{e}$ are shown in Fig.~2, using parameters for $^6$Li.  When the effective two-photon detuning $\delta$ is exactly zero, there is a minimum of the loss. If the decay rate between state $\ket {g_{1}}$ and $\ket{g_2}$ is negligible, as is the case in our scheme, the imaginary part of eq.~(\ref{scatteringlength}) is zero, as for EIT,  where the loss is completely suppressed. For these conditions, the adiabatic solutions for the molecular amplitudes $b_2$ and $b_e$ in terms of $b_1$ are
 \begin{align}
  b_{e}&=0,  \tag{10a}\label{8a}\\
 b_2&=-\frac{\Omega_{1}}{\Omega_{2}} b_1 \tag{10b}.
 \end{align}
 The bound-state probability $|b_1|^2$ can be obtained using the normalization condition, and is related to the molecular fraction measured in Ref.~\cite{Hulet}. Eq.~\ref{8a} clearly shows that there is no population for the excited molecular state $\ket e$: Atom pairs are completely trapped in the two ground molecular states, which is a consequence of quantum  interference, i.e., a ``dark'' state.

\begin{figure}
\includegraphics[width=3.0 in]{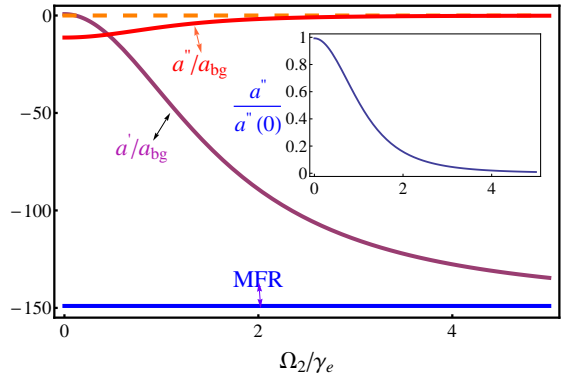}
\caption[example]
   { \label{fig:fig3}
Real $a^{'}/a_{bg}$ and imaginary $a^{''}/a_{bg}$ components of the scattering length as a function of  $\Omega_{2}/\gamma_{e}$ for $\Omega_{1}= 5\,\gamma_{e}$, and $\delta=0.05\,\gamma_{e}$. All other parameters are the same as in Fig. 2: The solid blue line is the scattering length without the laser fields (magnetic Feshbach resonance);  The dashed orange line  denotes $a''=0$. Inset: Loss ratio between the ``dark-state'' scheme and a typical single laser scheme (where $\Omega_{2}=0$) as a function of $\Omega_{2}/\gamma_{e}$.  }
\end{figure}

 The dark-state method enables control of the scattering length with very small loss by changing the Rabi frequencies $\Omega_{1}$
and $\Omega_{2}$ for a fixed magnetic field. Fig.~\ref{fig:fig3} shows $a^{'}$ and $a^{''}$ in units of the background scattering length $a_{bg}$ as a function of  the Rabi frequency $\Omega_{2}$ at $\omega_2=\omega_{eg_2}$, two-photon detuning $\delta=0.05\, \gamma_{e}$ and $\Omega_{1}= 5\,\gamma_{e}$. The scattering length dramatically changes by $\simeq 150 a_{bg}$ as the  Rabi frequency $\Omega_{2}$ is increased. The scattering length also can be made positive or negative, depending on whether the initial value of magnetic field is set to the BEC side (below) or the BCS side (above) the Feshbach resonance. The inset shows the loss ratio $a^{''}(\Omega_{2})/[a^{''}(\Omega_{2}=0)]$   between the ``dark-state'' scheme and typical single laser driving scheme ($\Omega_{2}=0$) as a function of $\Omega_{2}/\gamma_{e}$, demonstrating that the loss can be greatly suppressed using  ``dark'' states in the closed channel, compared with previous methods, where a control field with only one frequency is used~\cite{rempe}. For example, using the dark-state method with $\Omega_{2}=3\,\gamma_e$, the loss is two orders of magnitude smaller than the single field method of Ref.~\cite{rempe}. In addition, by dynamically changing $\Omega_{2}$ and $\Omega_{1}$, stimulated Raman adiabatic passage (STIRAP), can be used for coherent transfer of the populations between  $g_1$ and $g_2$, which is very important in the formation of the ro-vibrational ground molecules.

\begin{figure}[htb]
\includegraphics[width=8.8cm,height=4cm]{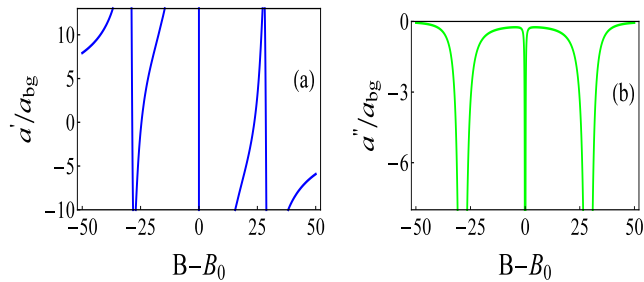}
\caption[example]
   { \label{fig:fig4}
Scattering length as a function of  $B-B_{0}$ for fixed laser parameters $\Omega_{1}=8\gamma_{e}$,
$\Omega_{2}=12\gamma_{e}$, $\omega_1=\omega_{eg_1}$, $\omega_2=\omega_{eg_2}$. All other parameters are the same as in Fig. 2: (a) $a^{'}/a_{bg}$ (b) $a^{''}/a_{bg}$.}
\end{figure}

The dark state method also enables control the magnetic field dependence of the scattering length for fixed laser parameters, as shown in Fig.~\ref{fig:fig4} where the real and imaginary parts of the scattering length,  $a^{'}/a_{bg}$ and $a^{''}/a_{bg}$,  are plotted as a function of  $B-B_{0}$. The plot shows a three peak structure, with a very narrow central resonance, characteristic of the dark-state method, which permits large changes in the scattering length for small changes in $B$. In contrast, using a single field control method~\cite{rempe}, there are two broad resonances and corresponding losses arising from typical Autler-Townes (AT) doublets.  The loss shows three peaks, similar to an EIT medium inside an optical cavity, where there are three transmission peaks: one narrow central peak corresponding to the ``dark'' state and two side peaks corresponding to dressed Rabi splittings~\cite{wu}.

The dark state method is readily implemented in fermionic $^6$Li, where $\ket{g_1}$ corresponds to the highest lying 38$^{th}$ vibrational state, located $1.58$ GHz below the singlet continuum of the ground electronic ($X^{1}\sum_{g}^{+}$) state. In this case, we can take $\ket{g_2}$ to be 37$^{th}$ vibrational state, located $53.5$ GHz below $\ket{g_1}$. The optical frequencies can be generated by frequency offset locking two diode lasers to a cavity. For comparison, in bosonic $^{85}$Rb, the two highest lying  vibrational states are located $195$ MHz and $1.53$ GHz below the singlet continuum. In this case, both frequencies can be generated by modulation of a single laser source.  For the excited electronic state   of $^6$Li ($A^{1}\sum_{u}^{+}$), we take $\ket {e}$ to be the $v^{'}=68$  vibrational state, which has the largest Franck-Condon factor with $\ket{g_{1}}$~\cite{Hulet,CoteJMS1999}. The transition wavelength is 673.7 nm, compared to 671.0 nm for  the atomic line. The Rabi frequency for the molecular transition $g_1\rightarrow e$ is $\Omega_1\simeq 0.59\,\text{MHz}\sqrt{I(mW/cm^2)}$~\cite{Hulet,CoteJMS1999}, and the spontaneous decay rate is  $\gamma_e\simeq 12$ MHz. Finally, the background scattering length is $a_{bg}=-1405\,a_0$~\cite{GrimmBroadFB}.

In the conclusion, we have shown  that the scattering length near a magnetic Feshbach resonance can be widely controlled and manipulated using a ``dark'' state optical method, when there is more than one  molecular state in the closed channel. In contrast to previous single optical field  methods, the closed-channel dark-state approach employs destructive quantum interference, arising from two closely-spaced optical transitions, to greatly suppress the two-body light-scattering induced loss and heating of the atomic gas.  The method has  important applications in ultracold quantum gases,  enabling rapid temporal and high resolution spatial control of interactions and studies of nonequilibrium dynamics on fast time scales, as well as studies of the dynamics unstable systems, such as Bose gases near Feshbach resonances and three-state Fermi gases.

This research is supported by the Physics divisions of the Air Force Office Office of Sponsored Research, the National Science Foundation, and the Army Research Office, and the Division of Materials Science and Engineering,  the Office of Basic Energy Sciences, Office of Science, U.S. Department of Energy.


\end{document}